\documentclass[twocolumn,pre,floats,aps,amsmath,amssymb,showkeys]{revtex4}
\usepackage{graphicx}
\usepackage{epstopdf}
\usepackage{color}

\DeclareGraphicsExtensions{.pdf,.eps,.png,.jpg,.mps}

\begin{document}


\title{Direct observation of chimera-like states in a ring of coupled electronic self-oscillators}

\author{L.Q. English$^1$, A. Zampetaki$^2$, P.G. Kevrekidis$^3$, K. Skowronski$^1$, C.B. Fritz$^1$, Saidou Abdoulkary$^4$}
\affiliation{$^1$Department of Physics and Astronomy, Dickinson College, Carlisle, PA 17013 \\
  $^{2}$ Zentrum f\"ur Optische Quantentechnologien,
Universit\"at Hamburg, Luruper Chaussee 149, 22761 Hamburg, Germany \\
$^3$Department of Mathematics and Statistics, University of Massachusetts, Amherst, MA 01003 \\
$^4$D\'epartement des Sciences Fondamentales, IMIP University of Maroua, P.O. Box 46, Maroua,  Cameroon
}

\begin{abstract}
  Chimera states are characterized by the symmetry-breaking coexistence of synchronized and incoherent groups of oscillators in certain chains of identical oscillators.  We report on the direct experimental observation of states
  reminiscent of such chimeras within a ring of coupled electronic (Wien-bridge) oscillators, and compare these to numerical simulations of a
  theoretically derived model.
  Following up on earlier work characterizing the pairwise interaction
  of Wien-bridge oscillators by Kuramoto-Sakaguchi phase dynamics,
  we develop a lattice model for a chain thereof, featuring an
  {\it exponentially decaying} spatial kernel.
  We find that for certain values of the Sakaguchi parameter $\alpha$, chimera-like states involving the coexistence of two clearly-separated
  regions of distinct
  dynamical behavior can establish themselves in the ring lattice, characterized by both traveling and stationary coexistence domains of synchronization. 
\end{abstract}



\maketitle

\section{Introduction}
Chimera states simultaneously featuring stable groups of synchronized and incoherent oscillators in coupled, spatially-extended systems have attracted enormous attention ever since they were first seen more than a decade
ago~\cite{kura,strogatz}. This is likely due to the counter-intuitive symmetry breaking that such spatio-temporal patterns display. Kuramoto and Battogtokh first discovered the possibility of such symmetry-broken states within a network of identical phase oscillators in numerical simulations of the complex Ginzburg-Landau equation with non-local coupling~\cite{kura}. Soon thereafter, Strogatz and Abrams performed a rigorous analysis and also coined the term ``chimera'' for
this novel state~\cite{strogatz,abrams}.

In the years since this analysis, a diverse array of
studies focusing on the numerical computation of chimeras revealed
the wide applicability and  robustness of such a feature (even against small spatial inhomogeneities) across many coupled-oscillator systems.
Furthermore,  distinct dynamical variants to Kuramoto's original chimeras
were discovered (transient vs. permanent, traveling vs.
standing, etc.)~\cite{knobloch1,carlo, knobloch2}. A  classification
of the different known chimera-like states can be found in the recent
contribution of Ref.~\cite{krischer}; see also references therein, as well as in Ref.~\cite{panaggio},
for a broad range of relevant examples. 

On the experimental side, the number of available paradigms is far
more limited. Nevertheless, recently chimera states have begun to be reported in mechanical \cite{martens}, chemical \cite{tinsley}, electro-chemical \cite{schmidt}, and opto-electronic \cite{hart} oscillator systems. An optical study also found chimera dynamics within liquid-crystal cells \cite{hager} where the coupling was
accomplished via computer feedback. Similar patterns were also seen in an electronic FM oscillator with time-delayed feedback \cite{larger}. In the mechanical case, a group of identical metronomes on elastically coupled swings revealed the possibility for a chimera state consisting of one synchronized group and another turbulent one~\cite{martens,haug}. The observation of near-harmonic chemical oscillations in the oxide-layer thickness of electro-oxidized silicon \cite{schmidt} found two-dimensional quasi-chimera states in which coherent and turbulent domains alternated in their
location \cite{haug}. A lattice of electronic oscillators mimicking neuron-like spiking was found to be
divided into quiescent and synchronized domains in~\cite{gambuzza}.

In this paper we report direct experimental observations of chimera-like states in a one-dimensional lattice of phase oscillators. We also develop
from first principles a corresponding theoretical model and perform numerical simulations of the equations approximately describing the experimental system and discuss the results. The lattice consists of 32 Wien-Bridge oscillators, each one bi-directionally and resistively coupled to its two nearest neighbors in a ring formation. This system has been the subject of recent synchronization studies \cite{temir,english} and shown to obey the Kuramoto-Sakaguchi phase oscillator
model \cite{sakaguchi} to a good approximation. Crucially, its dynamics can be fully captured (in a distributed, spatio-temporal way)
and characterized experimentally. 

While our numerical observations do not reveal turbulent or chaotic
states, we dub the observed features chimera-like states in the
following generalized sense: our system supports a {\it traveling} domain of nearly phase-locked oscillators inside of a {\it standing}
nearly out-of-phase background. Experimentally, we also see hints of these traveling chimera states (at the high end of the explored parameter window discussed
below), but in addition (for slightly lower parameter values) we also see stationary patterns of coexisting domains.
These also include the possibility of one of the co-existing states being
apparently chaotic.
In the simulations, such stationary domains could only be achieved by the introduction of impurities.  
More broadly, both experimental and numerical results suggest that the existence of  these chimera-like states relies on the presence of bi-directional coupling, as well as on the Sakaguchi-Kuramoto phase-delay parameter, $\alpha$, set to a value within a fairly narrow window.

Our presentation will be structured as follows. We will start by
describing the experimental setup and design (section II). We will then
present an approximate theoretical model emulating the
relevant electronic system (section III). In section IV, we will
give a series of simulations for guidance on the kind of
phenomenology that the model may exhibit. Finally, in section V
we will present our experimental results, and in section VI
we will summarize our findings and present some conclusions
and future challenges.

\section{Experimental setup} 
\begin{figure}
\centering
\includegraphics[width=3.5in]{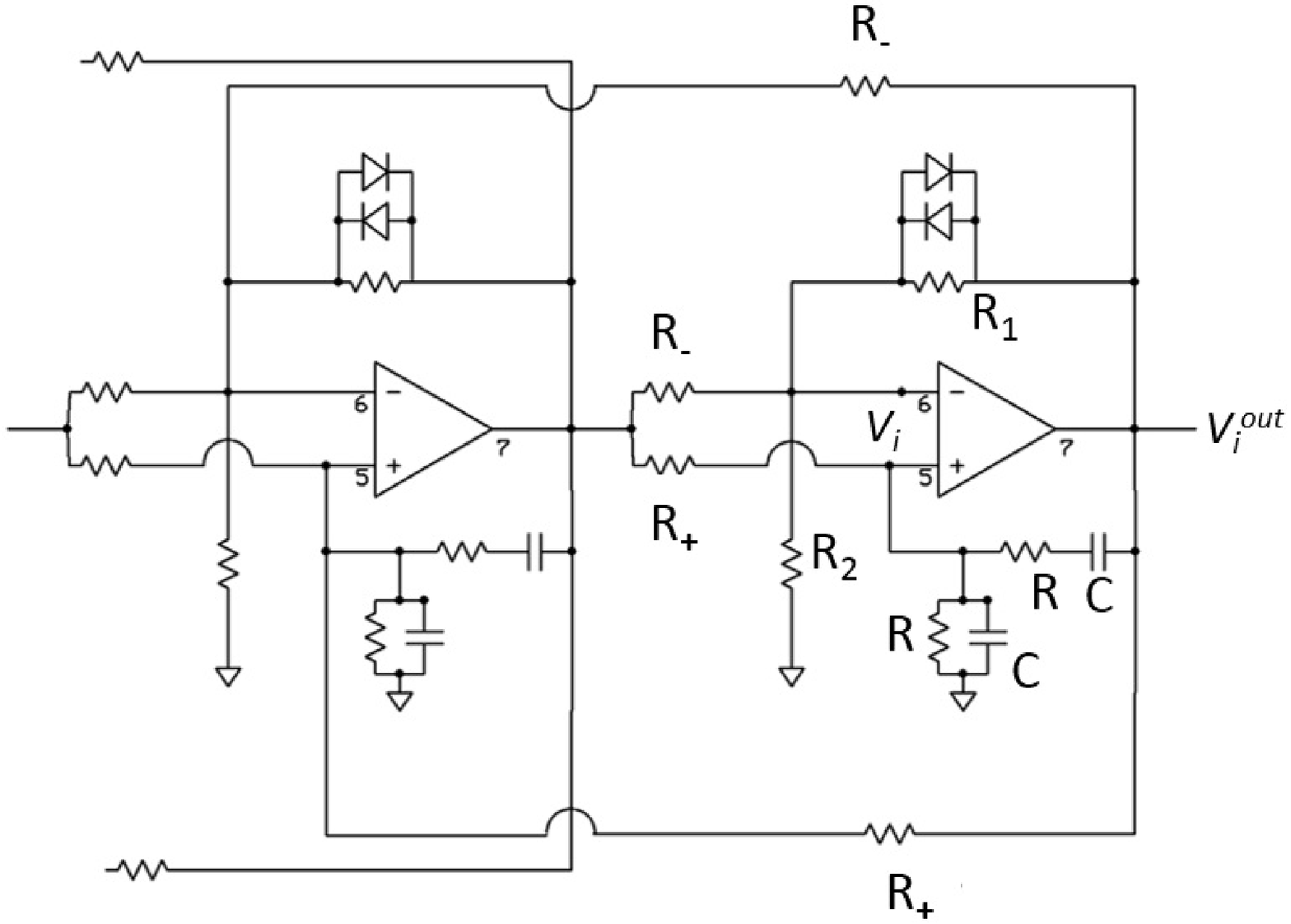}
\includegraphics[width=1.5in]{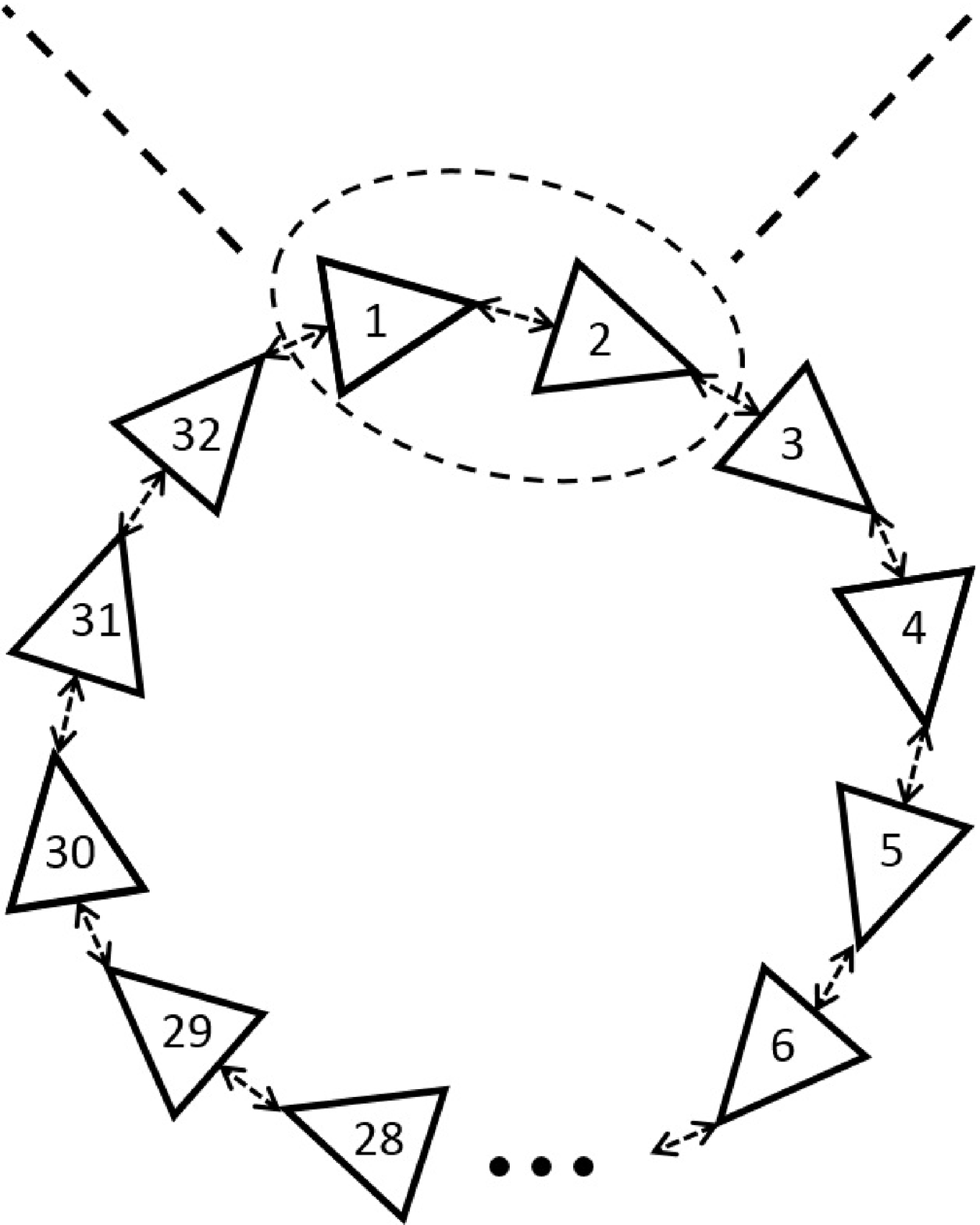}
\caption{Schematic of the coupled oscillator ring circuit with two Wien-bridge oscillators shown. The coupling is accomplished via two resistors to each of the two neighboring oscillators in the ring: from the output of the oscillator's op-amp to the non-inverting input of the two neighbors, via $R_+$ as well as to their inverting inputs, via $R_-$. Thus, the output of each op-amp is connected to both inputs of the two neighboring op-amps via $R_+ = R_-=$ 62 k$\Omega$ resistors.}
\label{circuit}
\end{figure}

Figure \ref{circuit} schematically depicts the experimental system. We couple 32 individual Wien-bridge oscillators in a ring topology. The oscillators have a natural frequency centered around 279.5 Hz with a standard deviation of 0.35 Hz. This means that the spread in natural frequency is only about 0.13 percent and
hence for practical purposes the oscillators are nearly identical in this measure. The coupling between nearest neighbors in the ring is accomplished resistively, and it can be either uni-directional or bi-directional in nature; shown in the figure is the latter configuration with two oscillators fleshed
out in the top panel. All oscillators are monitored simultaneously by a 32-channel digitizer. In previous studies \cite{temir, english}, only one resistor from the op-amp output of the first oscillator to the non-inverting input of the second (labeled $R_+$) was used to establish the (uni-directional) connection. The resulting oscillator phase interaction was found to be well
described by a Kuramoto-Sakaguchi model with a fairly low value of the
phase-delay parameter, $\alpha$. A second coupling channel between oscillators, via the resistor labeled $R_-$, now connects to the inverting input of the receiving op-amp, as shown in the diagram. 

\begin{figure}
\centering
\includegraphics[width=3.0in]{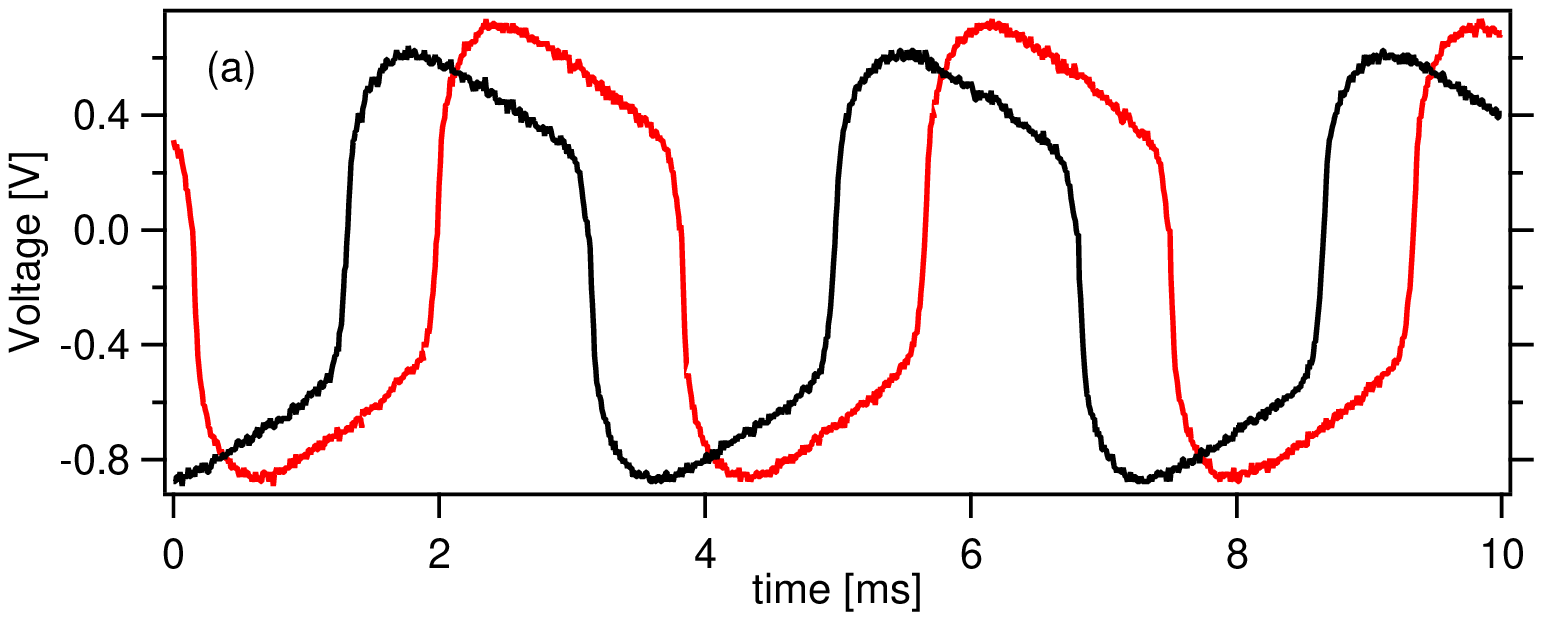}
\includegraphics[width=3.0in]{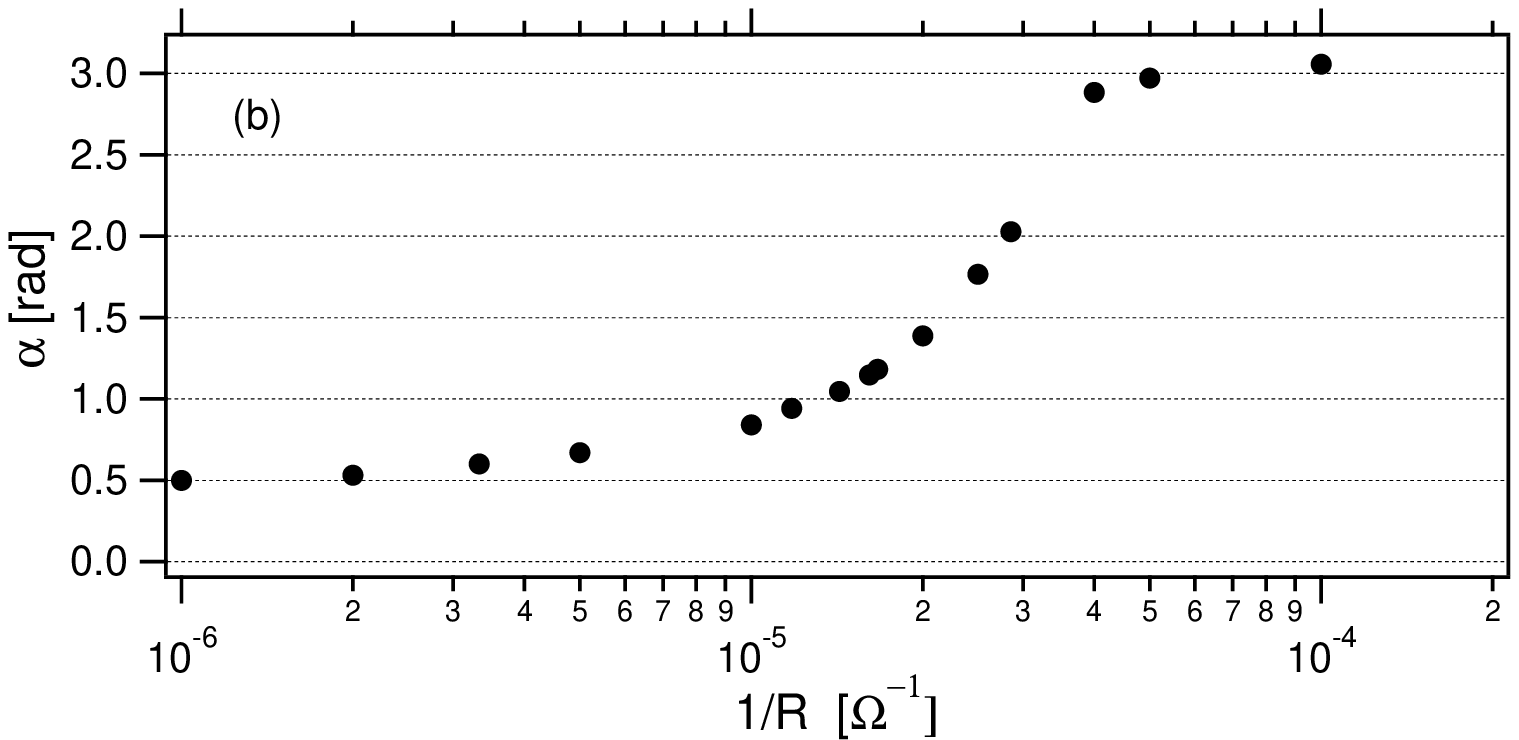}
\caption{The coupling scheme depicted in Fig.\ref{circuit} produces a Sakaguchi parameter, $\alpha$, that is somewhat below $\pi/2$ when the two coupling resistors have the value of 62 k$\Omega$. (a) One Wien-bridge oscillator (black trace) driving a second such oscillator (red trace), both of which have been fine-tuned to have the same natural frequency. We see that the driver is ahead in phase relative to the driven oscillator. This phase difference should be equal to $\alpha$ and is measured here as 1.2 rad. (b) When the inverting-input resistance, $R_-$ is adjusted while keeping $R_+$ constant, the Sakaguchi-parameter $\alpha$ can be continuously varied.}
\label{alpha_det}
\end{figure}

Since the effective value of $\alpha$ in the Kuramoto-Sakaguchi model plays an important role in the emergence of the chimera state \cite{strogatz}, it is essential to determine and control it experimentally. For the purpose of measuring $\alpha$, two Wien-bridge oscillators were first tuned to the same natural frequency (to within achievable precision) before adding two resistors, $R_-$ and $R_+$, that couple them uni-directionally. This oscillator pair should then be described by the system,
\begin{eqnarray}
\dot{\phi}_1 &=& \omega_1 \nonumber \\
\dot{\phi}_2 &=& \omega_1 + K \sin(\phi_1-\phi_2-\alpha).
\end{eqnarray}
If the two oscillators synchronize, $\dot{\phi}_1=\dot{\phi}_2$, and this forces $\phi_1-\phi_2$ to be equal to $\alpha$. Figure \ref{alpha_det}(a) shows this case for the symmetric coupling situation of $R_+ = R_- = 62 $k$\Omega$. As predicted, the driving oscillator (oscillator 1) leads the driven oscillator (oscillator 2) in phase. Measuring their phase difference then yields $\alpha$.  

It is also desirable to possess some experimental control of $\alpha$. Several modifications to the purely resistive coupling were tested for this purpose, such as incorporating coupling capacitors. However, simply adding the coupling resistor, $R_-$, proved to be the most efficient way to tune $\alpha$ over a large interval. The effect of this coupling resistor, $R_-$, is illustrated in Fig. \ref{alpha_det}(b). Note that here $R_+$ was fixed at 62 k$\Omega$. For very large values of $R_-$, the phase-delay parameter $\alpha$ is close to the value of 0.5 found previously, whereas for small values, $\alpha$ saturates at $\pi$. The most sensitive dependence is seen to occur within the resistance interval of 10 k$\Omega$ and 100 k$\Omega$. In the subsequent data-sets, we choose $R_- = R_+ = 62$ k$\Omega$. This combination produces an $\alpha$-value of 1.2 rad.

\section{The model}
Under certain assumptions, the governing equations for the voltages at each node of the lattice can be derived as in Ref.~\cite{english2}, by starting from Kirchhoff's node rule at the positive and negative Op-amp inputs of the $i^{th}$ oscillator. Relegating the details to the appendix, an intermediate result
for the temporal evolution of the voltage
at the $i$-th node takes the following form:
\begin{align}
&V_i^{\prime\prime}+\left(2-\frac{R_1}{R_2}-\frac{2R_1}{R_-}+\frac{2R}{R_+}\right)V_i^{\prime} + \left(1+\frac{2R}{R_+}\right)V_i -\frac{R}{R_+} \nonumber \\
&\left(V_{i+1}^{out}+V_{i-1}^{out}\right)+\left(\frac{R_1}{R_-}-\frac{R}{R_+}\right)\left({({V}_{i+1}^{out})}^{\prime}+{({V}_{i-1}^{out})}^{\prime}\right)=0.
\label{eq1}
\end{align}
Here, the prime denotes differentiation with respect to non-dimensional time $\tau = t/(RC)$. $R_1$ and $R_2$ are the two resistors of the non-inverting amplifier part of each Wien-bridge circuit, with $R_1$ taken to have a nonlinear voltage-dependence (due to the two diodes in parallel with it). The resistor $R$ is in the filter circuit from Op-amp output to positive (non-inverting) input, and $R_+$ and $R_-$ denote the two coupling resistors; for the details
of these quantities in the coupled oscillator ring circuit,
see Fig.~\ref{circuit}.

Note that $V_i$ is the voltage of node $i$ at the input of the Op-amp, whereas $V_i^{out}$ is the voltage at the output of the Op-amp. Thus, Eq.~(\ref{eq1}) is not yet sufficient in order to solve for the voltages; we must also find a relationship between the input and output voltages. By examining current flow into or out of the node at voltage $V_i$ at the inverting input, we can derive the following relationship:
\begin{equation}
\left(1+\frac{R_1}{R_2}+\frac{2R_1}{R_-}\right) V_i = V_i^{out} + \frac{R_1}{R_-}(V_{i+1}^{out}+V_{i-1}^{out}).
\end{equation}
We can write this in matrix form as,
\begin{equation}
{\bf V}^{out} = c \mbox{ } \mathcal{A}^{-1}\mbox{ } {\bf V}^{in},
\label{eq2}
\end{equation}
where $c=1+\frac{R_1}{R_2}+\frac{2R_1}{R_-}$, and 
\[ \mathcal{A} = \left[ {\begin{array}{cccccc} 1 & \gamma & 0 & ... & & \gamma \\ \gamma & 1 & \gamma & 0 &  &... \\ 0 & \gamma & 1 & \gamma & 0 &... \\& & &... \\ \gamma & 0&... & &\gamma & 1 \end{array} } \right] \] 
with $\gamma=\frac{R_1}{R_-}$.
$\mathcal{A}$ is a tri-diagonal N x N matrix encapsulating the novel nearest-neighbor coupling scheme. If no coupling exists into the negative (inverting) input of the Op-amps, then $\gamma=0$ and $\mathcal{A}$ reduces to the identity matrix. 

Equation (\ref{eq2}) can also be adapted to the derivatives of the voltages appearing in Eq.~(\ref{eq1}). Note, however, that in the experiment $R_1$ is voltage-dependent, and thus implicitly time-dependent. To make the modeling somewhat more tractable, let us consider the slightly modified Wien-bridge oscillator where the nonlinearity is contained in $R_2$, and where $R_1$ is constant. For concreteness, we assume that,
\begin{equation}
R_2=R_{20}(1+b V_i^2).
\label{nonlin}
\end{equation}
Given the symmetric nature of the configuration, we expect this
voltage dependence to be generically valid for small voltages.
Then, from Eq.~(\ref{eq2}) we obtain:
\begin{equation}
\dot{V}_i^{out} = c \mbox{ } \sum_j \mathcal{A}_{ij}^{-1}\mbox{ } \dot{V}_j-\frac{R_1}{R_2^2}\frac{dR_2}{dV_i}\left(\sum_j \mathcal{A}_{ij}^{-1}\mbox{ } V_j\right)\mbox{ } \dot{V}_i
\label{eq2dot}
\end{equation}

Equations (\ref{eq2}) - (\ref{eq2dot}) together with Eq.~(\ref{eq1}) represent the governing equations of motion for the voltages on this lattice.
A key observation is that the inverse of the tri-diagnonal matrix $\mathcal{A}$ is {\it not} itself tri-diagonal, but rather
involves a form of non-local coupling, although it will be heavily diagonal-centered for small $\gamma$. Consequently, the system of Eq.~(\ref{eq1}) is, in fact, defined by lattice-node coupling that goes {\it beyond nearest-neighbor}
and hence bears a prototypical feature that is often associated with
chimera-bearing states, namely a non-local coupling. 
There exist analytic solutions for the inverse of symmetric tri-diagonal matrices like $\mathcal{A}$. In general, when the diagonal entries are all equal to $a$, and the nearest off-diagonal entries equal to $\beta$, then for $i>j$, we have~\cite{fonseka},
\begin{equation}
\nonumber
[A^{-1}]_{i,j} = (-1)^{i+j} \frac{1}{\beta} \frac{U_{j-1}(a/2\beta) U_{n-i}(a/2\beta)}{U_n(a/2\beta)},
\end{equation}
where $U$ are the Chebyshev polynomials of the form,
\begin{equation}
\nonumber
U_n(x)=\frac{\sinh(n+1)\theta}{\sinh \theta}, \mbox{ and  } \cosh \theta = x.
\end{equation}

If we adapt this result to our matrix, we obtain the following formula for the entry in the first row of the inverse:
\begin{equation}
[A^{-1}]_{x,1} = (-1)^{x+1} \left(\frac{1}{\gamma}\right) \frac{\sinh \kappa (N+1-x)}{\sinh \kappa (N+1)},
\label{coupling}
\end{equation}
where $x$ is an integer between 1 and $N$, and $\kappa = \cosh^{-1}\left(\frac{1}{2\gamma}\right)$. Using the approximation that $\sinh \kappa(N+1) \approx \exp[\kappa (N+1)]/2$, which should hold very well for large lattices (of sufficiently large size N), we arrive at:
\begin{equation}
[A^{-1}]_{x,1} \cong (-1)^{x+1} \left(\frac{1}{\gamma}\right) e^{-\kappa x}.
\end{equation}
Thus, we obtain a spatial kernel that is essentially exponentially decaying with a decay constant that depends only on $\gamma$: the larger the value of $\gamma$, the smaller is the decay constant. Note also the alternating sign in the kernel. 
To illustrate this point with a concrete example, assume that $N=100$ and that $\gamma=0.1$. This yields a decay constant of $\kappa = 2.3$, and the first row of $\mathcal{A}^{-1}$ reads: \{$1.02, -0.103, 1.04*10^{-2}, -1.05*10^{-3}, 1.06*10^{-4}, 1.07*10^{-5}, ...,-0.103$\}. This means that $V_{1}^{out}$ will be predominantly given by $V_{1}$, but $V_0$ and $V_2$ will still also play a role, with a progressively small contribution from $V_{99}$ and $V_3$ and beyond.
In the experimental system, we have $R_1=27 k\Omega$ and $R_{-}=62 k\Omega$, so $\gamma=0.435$, which leads to a spatial decay constant $\kappa=0.54$.
This results in a weaker decay of the relevant kernel and a broader
range inter-neighbor coupling than the example above. In any event,
in our numerical computations that will follow, we consider all
neighbors with the respective pairwise coupling as dictated
by Eq.~(\ref{coupling}).

\section{Numerical Simulations}

In order to motivate the experimental results that will follow,
we have performed numerical simulations of the aforementioned theoretical model, starting with random initial conditions (voltage values $V_i(0)$ in the interval $[-1,1]$ and  $V_i'(0)=0$).
By adjusting properly the values of the effective parameter $b$ and the gain resistances $R_1$ and $R_2$ we can obtain stable oscillations in the long time dynamics over a large range of $R_{-}$ values. Keeping $R_+$ fixed
at 62 k$\Omega$ and incrementing $R_{-}$, we observe that the system of
bi-directionally coupled oscillators passes from an anti-synchronized phase
to a fully synchronized one.

\begin{figure}[htbp]
\begin{center}
\includegraphics[width=8.6cm]{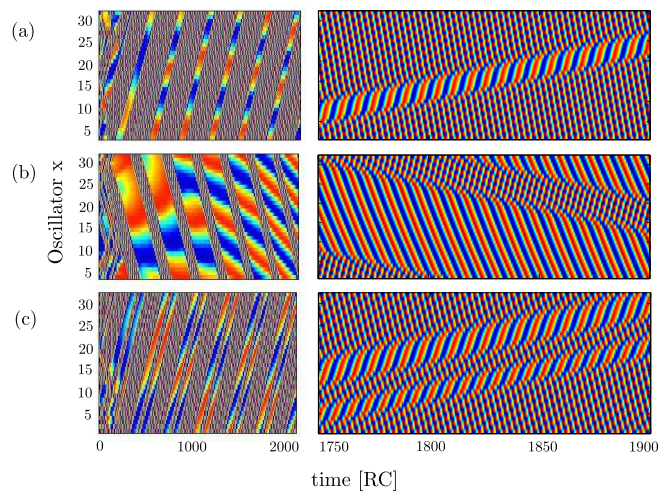}
\end{center}
\caption{\label{tra_ch} Three different cases (a), (b), (c) of simulation results derived by the aforementioned theoretical model (Eqs.~(\ref{eq2})-(\ref{eq2dot}) together with Eq.~(\ref{eq1})) exhibiting 
  a chimera-like behavior. Displayed are the simulated voltages (depicted
  by the color) as a function of time (x-axis) and oscillator index (y-axis).
The left panels account for the dynamics of the system in a large time interval whereas the right panels are zooms of the long-time behavior. Obviously there is a traveling chimera-like in all three
cases with a different width and number of the synchronized domains. In all cases $R_{+}=62 k\Omega, R_1=9k\Omega, R_2=2.7k\Omega, b=5$, whereas for
(a) $R_{-}=100k\Omega, R=4.7k\Omega$, (b) $R_{-}=118k\Omega, R=4.7k\Omega$, (c) $R_{-}=118k\Omega, R=4k\Omega$.}
\end{figure}

In the intermediate regime the dynamics of the oscillator ring is prone to the formation of chimera-like patterns consisting of mixtures of
the two distinct phases (synchronized and anti-synchronized
ones). As shown in Fig.~\ref{tra_ch}, these typically have the form of traveling patterns of a nearly in-phase domain within a nearly anti-phase background, or vice versa. These traveling synchronization domains are seen to persist in time. The particular traveling chimera pattern can consist of a narrower (Fig. \ref{tra_ch} (a)) a wider (Fig. \ref{tra_ch} (b)) or even multiple (Fig. \ref{tra_ch} (c)) synchronized domains
with both positive (Figs. \ref{tra_ch} (a),(c)) and negative (Figs. \ref{tra_ch} (b)) velocities. Which pattern is finally realized depends on the exact parameters 
as well as on the initial conditions. Irrespectively of the latter, however, the tendency of the model to form traveling chimera-like
patterns in a specific parameter regime is clear. However, it should
be highlighted that we label these states ``chimera-like'' (rather than
genuine chimeras) because they bear the characteristic of co-existence
of (in fact) moving domains of Wien-bridge oscillators, but
neither of the distinct phases appears to be turbulent or decoherent in its nature
within our numerical computations. However, it is relevant to mention
that the latter ``chaotic'' characteristic will be seen to arise
in our experiments below (and its absence in the presented
computations may be likely
linked to the limitations of the model).

\begin{figure}[htbp]
\begin{center}
\includegraphics[width=8.6cm]{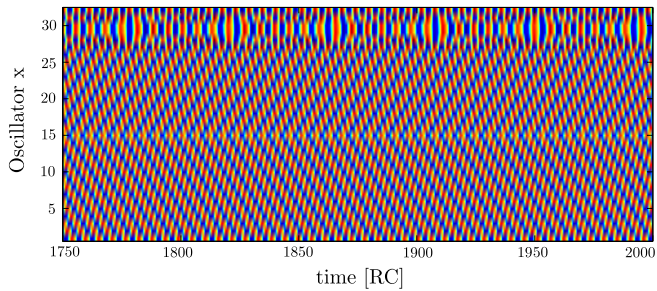}
\end{center}
\caption{\label{imp_ch} Simulated voltages (displayed as color) as a function of time (x-axis) and oscillator index (y-axis) for the case of Fig. \ref{tra_ch} (a) but with an impurity.
The impurity is due to the value of the resistance for the 15th oscillator being $R=3.5k\Omega$, whereas for all other oscillators $R=4.7 k\Omega$.}
\end{figure}

Given that our theoretical model favors the formation of traveling, rather than stationary chimeras, there is a question about whether the presence of an impurity can lock
these chimera patterns in space. For this reason, we performed further numerical simulations in which the value of the resistance $R$ of a certain oscillator differs from that of the rest  in the ring. 
It is found that for a strong enough impurity (a difference of about 25\%)  a stationary pattern can be stabilized, as shown in Fig. \ref{imp_ch}. Here small intermittent synchronized domains appear inside an anti-synchronized phase.
Furthermore, the defect in this case can be classified, in accordance
with the classification of~\cite{bjorn}, as a sink (given its concurrent
emission of wavetrains on the two sides).

\section{Experimental results and Discussion}

\begin{figure}
\centering
\includegraphics[width=3.4in]{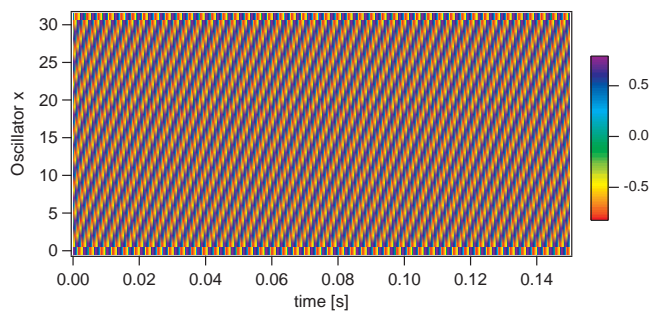}
\caption{The measured voltages (displayed as color) as a function of time (x-axis) and oscillator index/position (y-axis). The connections are all uni-directional (clockwise around the ring). The oscillators quickly synchronize into a perfectly phase-locked mode that obeys the periodic boundary condition. No oscillators are drifting in this state.}
\label{unidirec}
\end{figure}

Motivated by the above computational illustrations of the potential
for co-existing phases within the model, we now turn to the main
core of the present contribution, consisting of experimental investigations
of the potential formation of chimera-like states in this system.
We once again caution the reader that 
the experimental system is only {\it approximately} described by the model, and hence the observed deviations from the simulation results may be attributed to model limitations. The main difference is (a) that we operate the experimental Wien-bridge oscillators at a higher gain value of $g=1+R_{10}/R_{20}=10$, and (b) that the diode-pair is in fact lowering the effective resistance of $R_1$ and not increasing the resistance of $R_2$, and (c) that higher order terms would be needed in Eq.~(\ref{nonlin}) to accurately model the effect of diodes.
This latter feature is worthwhile of additional independent study.
Nonetheless, the model is phenomenologically consistent with the gain
saturation for larger voltages caused by the diodes and already features
a number of intriguing phenomenological features motivating the
presented experiments.
   
Having stated these caveats, let us turn to the experimental picture. We start the investigation with near-identical oscillators coupled in the {\it uni-directional} ring topology. Here, the oscillators synchronize themselves into a perfectly phase-locked mode that obeys the periodic boundary condition of $\Delta \phi = 2\pi m / N$. This is illustrated in Fig.~\ref{unidirec}, which depicts a mode number of $m=9$, resulting in a phase advance of $\Delta \phi = 1.76$ rad. Here the coupling resistors, $R_{-}$ and $R_{+}$, are both at 62 k$\Omega$. Such forward-facing rolls of synchronization are very stable states for this coupling configuration. Since each oscillator is effectively driven by the previous one without acting back on it, such patterns are naturally expected.

\begin{figure}
\centering
\includegraphics[width=3.55in]{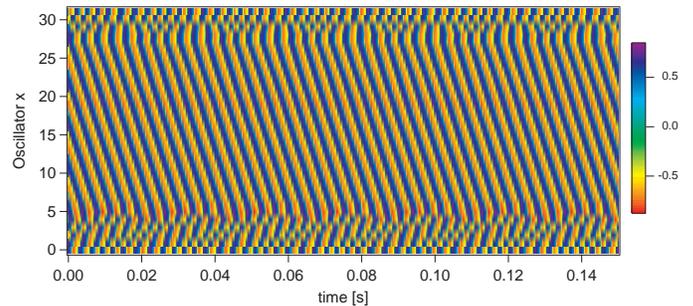}
\caption{The measured voltages (displayed as color) as a function of time (x-axis) and oscillator index/position (y-axis). Here, the connections are bi-directional. We clearly see the synchronized cluster between oscillators 5 and 28 surrounded on both sides by drifting oscillators. This
  represents the observation of a chimera state.}
\label{result1}
\end{figure}
When additional coupling resistors of the same value (62 k$\Omega$) are added to make the ring bi-directional, however, the dynamics changes significantly. In this situation, due to the underlying symmetry one might expect a state of perfect synchrony, corresponding to a mode number of $m=0$, but this is not what is observed. Figure \ref{result1} depicts a typical data-set.
A phase-locked cluster of oscillators can clearly be discerned. This cluster extends roughly between oscillator 5 and 28. Outside of this cluster, no stable phase pattern is evident. 
The spatial location of the synchronized cluster is likely determined by small impurities in the ring lattice (both on-site and in the coupling). We have made sure that the uncoupled frequencies of all 32 oscillators are within a very narrow band; the distribution mean and standard deviation are given by 279.5 Hz and 0.35 Hz, respectively. However, this does not imply that all the $R$ and $C$ values are this narrowly distributed separately. A more likely source of inhomogeneity enters via the gain resistors, $R_1$ and $R_2$, all of which have a tolerance of 5$\%$.

\begin{figure}
\centering
\includegraphics[width=3.3in]{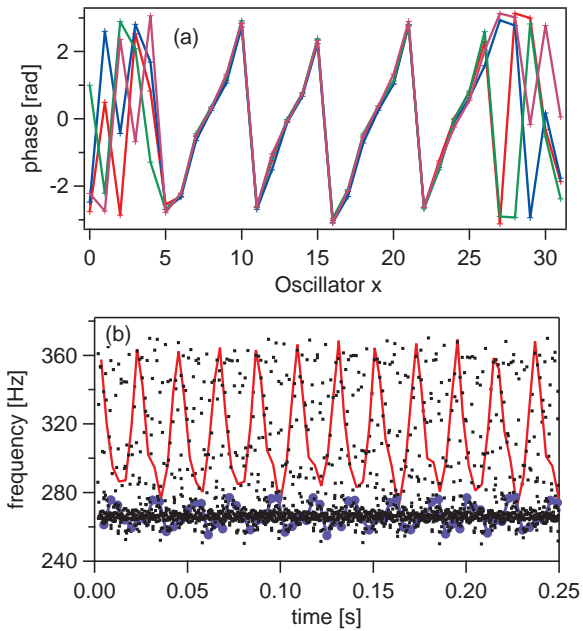}
\caption{Characterization of the measured chimera-like state. (a) The phase of each oscillator at the following snapshots in time: t=0.280, 0.284, 0.288,
  0.292, 0.332s. We see that the oscillators within the synchronized cluster maintain a fixed phase relationship with one another, whereas the background oscillators assume random phases at these times. (b) The frequencies as a function of time of all oscillators (depicted as dots), as computed from the experimental data. The cluster oscillators have a constant frequency of $\Omega=265$ Hz, whereas the frequencies of the decoherent oscillators fluctuate substantially with time while also staying above $\Omega$. The detailed frequency evolution is depicted for two particular oscillators as lines. The red line shows  oscillator 3, whereas the blue line (with circle markers) follows oscillator 5. We see that the periodic interaction with the cluster causes regularly-spaced frequency pulling.}
\label{result2}
\end{figure}

In order to investigate the chimera-like pattern in this system more closely, let us first extract the phase information of the individual oscillators from their raw voltage data. This was accomplished using the well-known Hilbert transform (described, for instance, in Ref. \cite{temir}). Note that the phases obtained in this fashion do not advance at exactly constant speeds over a single period, as the voltage oscillations deviate substantially from sinusoids. Nonetheless, inspection of the phase-profiles indicates no abrupt or large-scale fluctuations in phase-velocity within a single period of oscillation, thus providing us with a useful measure of oscillator phase.

Figure \ref{result2}(a) shows the phase of each oscillator at five equally-spaced instants of time. It is evident that the oscillators within the synchronized cluster from x=5 to x=28 maintain a fixed phase relationship with one another, whereas the background oscillators assume fairly random phases at these times.  Furthermore, the synchronized cluster is characterized by a non-zero phase advance of roughly $\Delta \phi \cong \pm$ 1.2 rad. While Fig. \ref{result2}(a) does not show it, close inspection of Fig. \ref{result1} suggests that there may be, however, a discernible pattern in the decoherent part. There we can see a cluster of nearly in-phase oscillators traveling in what appears to be a domain of out-of-phase oscillators. The overall effect is that the instantaneous phases will appear scattered over all angles, but the underlying reason may be the
periodic emergence of ``bursts'' of traveling in-phase clusters.    

Let us now turn to the dynamic oscillator frequencies. A straightforward way of computing these frequencies uses the interpolated (upward) zero-crossings in the voltage time-series (as described in more detail in Ref. \cite{english}), and this produces discrete frequency values that are averaged over each period of oscillation. Figure \ref{result2}(b) plots these dynamic frequencies computed from the raw data in Fig. \ref{result1} as dots. The dots corresponding to oscillators within the synchronized cluster all collapse onto the horizontal line at around 265 Hz. This group is characterized by a very constant, stable frequency located well below the natural (isolated oscillator) frequency of 280 Hz. Oscillators in the unsynchronized group  have frequencies that fluctuate quite substantially over time, depending on their phase relationship with respect to the synchronized cluster. It is also evident that the average frequency in this group is substantially above the synchronization frequency. 

To examine these fluctuations more closely, the figure highlights the dynamic-frequency evolution of two particular oscillators, one located at $x=3$ (red trace) and the other at $x=5$ (blue trace). The oscillator at $x=3$ is seen to periodically slow down to nearly match $\Omega=265 Hz$, and then speed up to a maximum frequency of around 360 Hz, almost 100 Hz above $\Omega$. The oscillator at $x=5$ is the first oscillator in the cluster, and we see that its frequency (blue trace) oscillates only slightly about the cluster frequency, $\Omega$. 

\begin{figure}
\centering
\includegraphics[width=3.3in]{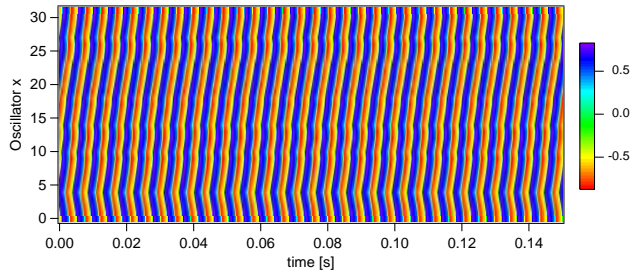}
\caption{After removing all $ R_{-} $ resistors, the effective phase-delay parameter, $ \alpha $, is lowered to around $0.5$ rad and the spatial kernel
  now involves only nearest neighbors. Here the chimera-like state is not observed. Instead, the oscillators are phase-locked and arrange themselves in near-vertical rolls.}
\label{lowalpha}
\end{figure}

To demonstrate that the chimera-like state relies on a sufficiently large value of the phase-delay parameter, $\alpha$, we remove all $R_{-}$ coupling resistors between oscillators, so that $\alpha$ reverts back to the lower value of around 0.5 rad. The $R_{+}$ resistors are still in place to provide bi-directional coupling. The results are shown in Fig. \ref{lowalpha}. We see that all oscillators are phase-locked with one another in a pattern of near vertical rolls. No oscillators are observed to drift in phase. Even if this pattern is temporarily perturbed via a momentary introduction of an impurity, it quickly reconstructs itself when the impurity is gone. In this $\alpha$ regime, a chimera-like state cannot be induced in the system.

\begin{figure}
\centering
\includegraphics[width=3.3in]{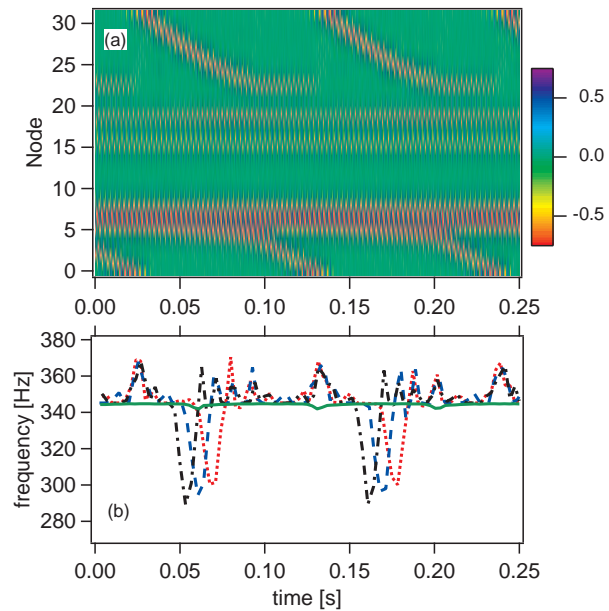}
\caption{(a) Spatial derivative of the temporal voltage-signal from the ring lattice of 32 oscillators. The solid (green) regions indicate perfect anti-phase synchrony, within which (chiefly) traveling nearly in-phase oscillatory patterns emerge. (b) The dynamic frequency of oscillator at nodes 10 (solid), 25 (dotted, red), 26 (dashed, blue) and 27 (black, dot-dashed) manifesting a substantial
  frequency drop as the traveling synchronized phase patterns pass by.}
\label{50_Ohm}
\end{figure}

We can also go higher in effective $\alpha$ by using a value for all negative coupling resistors of $R_{-}=50 k\Omega$. In this regime, the anti-phase domain is dominant with nearest neighbors oscillating exactly out-of-phase. However, smaller propagating islands of nearly in-phase oscillators are observed, reminiscent of the simulation results. Figure \ref{50_Ohm} (a) shows not the raw phase data, but instead the numerical spatial derivative, as approximated by
a centered difference, which accentuate the traveling patterns visually. In fact, the solid (green) areas here represent the regions of anti-synchronized behavior, whereas the traveling synchronized cluster can be seen quite clearly.
Interestingly, a synchronized nearly stationary region can also be detected
in the vicinity of oscillators $6$--$8$, which may be triggered by
an impurity.

In addition, we can examine the dynamic frequency of each oscillator, as before. The result is shown in Fig. \ref{50_Ohm}(b) for oscillator 10 (solid, green trace) as well as a group of three oscillators, 25 through 27 (dotted, dashed, and dot-dashed, respectively). We see that when this traveling cluster sweeps over a particular oscillator, the oscillator's dynamic frequency temporarily drops until the pattern has passed it by. Thus, the traveling domain is not only characterized by a much lower phase difference between neighboring oscillators;
it simultaneously manifests as a lower oscillation frequency.

Reference \cite{krischer} recently proposed a classification scheme for chimeras resting on spatial and temporal correlation parameters and their distributions. According to this scheme, in the simulations, we see {\it stationary moving} chimeras. The discrete second spatial derivative is approximately zero in the sync- and anti-sync domains, but at their boundaries, it is non-zero. In the experiment, we observe a similar type of chimera state for $R_-=50 k\Omega$. For $R_-=62 k\Omega$, however, we also observe something akin to {\it stationary static} chimeras (in accordance with the diagnostics and the classification
developed therein).

\section{Conclusions and Future Work}
We have found both experimentally and numerically that the ring-lattice of
bi-directionally coupled Wien-bridge oscillators with its exponentially
decaying spatial kernel admits chimera-like states.
On the contrary, uni-directionally coupled oscillators feature
uniformly phase-locked states.
We motivated the co-existence of different phases through a theoretical
model and its numerical simulation; in fact, the latter featured the
traveling of one or even multiple intervals of oscillators of one
type (in-phase) within a pattern of a different type (anti-phase).
The experimental measurements of a ring of 32 oscillators
that followed this qualitative theoretical/numerical investigation constitute,
to the best of our understanding,
one of the few direct observations of such a chimera-like state in an
extended system.
A crucial feature that may be responsible for such observations are
the non-local nature of the coupling of these oscillators that
was quantified herein.
In the experiment, we also find evidence of such traveling
synchronization domains (in qualitative agreement with our
numerical findings), however, the situation appears to be somewhat richer.
In particular, we also find non-traveling patterns of coexisting domains,
most likely pinned by slight inhomogeneities in oscillator properties.   

Naturally, the present work opens a significant array of new
directions. On the one hand, while here we examined some prototypical
parametric regimes of the model, it would be of interest to conduct
a deeper theoretical study of the possible outcomes of the
numerical experiments. On the other hand, it would be especially
useful to attempt to systematically characterize the dependence
within Eq.~(\ref{nonlin}) of the resistance on the voltage (i.e.,
the source of the model's nonlinearity) with the aim to produce
a model that more adequately captures the experimentally observed
phenomenology. Armed with these theoretical/numerical tools, one could
also embark in other directions associated with the potential that
this type of highly tractable, space-time resolved experiments offer.
In particular, in line with some of the most recent theoretical
investigations, one could explore the possibility of formation of chimera-like
states in higher dimensional setups, such as the twisted and spiral
chimera states of~\cite{knobl3}. Such studies will be deferred
to future investigations.

\begin{acknowledgments}
We thank David Mertens for some helpful early discussions about oscillator ring lattices. P.G.K. gratefully acknowledges
the support of NSF-PHY-1602994, as well as from the ERC under FP7, Marie Curie
Actions, People, International Research Staff Exchange
Scheme (IRSES-605096), the Alexander von Humboldt Foundation
and the Stavros Niarchos Foundation via the Greek Diaspora Fellowship
Program.
\end{acknowledgments}

\appendix
\section*{Appendix}

Here we give some of the steps that lead from the circuit equations to the system's governing equations of Eq.~(\ref{eq1}). We start by considering the currents going in and out of a node at the negative op-amp input of the $i^{th}$ oscillator. By the Kirchhoff node rule in the circuit of Fig.~\ref{circuit},
\begin{equation}
\frac{V_i}{R_2}=\frac{V_i^{out}-V_i}{R_1}+\frac{V_{i+1}^{out}-V_i}{R_{-}}+\frac{V_{i-1}^{out}-V_i}{R_{-}}.
\end{equation}
This then leads to,
\begin{equation}
\left(\frac{1}{R_1}+\frac{1}{R_2}+\frac{2}{R_{-}}\right)V_i=\frac{V_i^{out}}{R_1}+\frac{V_{i+1}^{out}+V_{i-1}^{out}}{R_{-}}.
\label{minus}
\end{equation}

Now, if we examine the positive op-amp input, we see that
\begin{equation}
V_i^{out}-V_i=V_{cap}+I_1 R,
\label{intermed}
\end{equation}
\noindent where $I_1$ is the current flowing through resistor $R$ in series with capacitor $C$, and $V_{cap}$ is the voltage across the capacitor. However, the current $I_1$ can be computed as follows:
\begin{equation}
I_1=C \dot{V}_i+\frac{V_i}{R}+\frac{1}{R_{+}}(V_i-V_{i+1}^{out}+V_i-V_{i-1}^{out}).
\label{current}
\end{equation}
Substituting Eq.~(\ref{current}) into Eq.~(\ref{intermed}) yields,
\begin{equation}
V_i^{out}=V_{cap}+(RC)\dot{V}_i+2(1+\frac{R}{R_+})V_i-\frac{R}{R_+}(V_{i+1}^{out}+V_{i-1}^{out}).
\label{plus}
\end{equation}	

Combining Eqs.~(\ref{minus}) and (\ref{plus}), we take another time derivative on both sides, substitute $C\dot{V}_{cap}=I_1$, and obtain after some further manipulation the following form:
\begin{align}
RC \ddot{V_i}+\left(2-\frac{R_1}{R_2}-\frac{2R_1}{R_-}+\frac{2R}{R_+}\right)\dot{V_i}+\left(\frac{1}{RC}+
\frac{2}{R_+C}\right)V_i \\ \nonumber
-\frac{1}{R_+C}(V_{i+1}^{out}+V_{i-1}^{out}) + \left(\frac{R_1}{R_-}-\frac{R}{R_+}\right)(\dot{V}_{i+1}^{out}+\dot{V}_{i-1}^{out})=0.
\end{align} 
Finally, introducing non-dimensional time, $\tau=\frac{t}{RC}$, we arrive at Eq.~(\ref{eq1}).

\end{document}